# PERMITTIVITY AND PERMEABILITY DETERMINATION FOR HIGH INDEX SPECIMENS USING PARTIALLY FILLED SHORTED RECTANGULAR WAVEGUIDES

Mark M. Scott,[1] Daniel L. Faircloth[2], Jeffrey A. Bean,[1] and Kenneth W. Allen[1]

[1] Advanced Concepts Laboratory, Georgia Tech Research Institute, Atlanta, GA 30318 USA; Corresponding author: kenneth.allen@gtri.gatech.edu

[2] IERUS Technologies, Inc., Huntsville, AL 35805 USA;

**ABSTRACT**: *A method for determining the permittivity and permeability for specimens with high refractive index and variable shape is investigated. The method extracts the permeability and permittivity tensor elements from reflection measurements made with a partially-filled shorted rectangular waveguide on an electrically small specimen. Measurements are performed for two isotropic, heavily loaded coaxial magnetic composites. Supporting measurements from a stripline cavity and coaxial airline are used to validate the method. The results demonstrate the methods ability to handle frequency dispersive and high index materials.* © 2015 Wiley Periodicals, Inc. Microwave Opt Technol Lett XX:XXXX–XXXX, 2015; View this article online at wileyonlinelibrary.com. DOI XX.XXX

1. INTRODUCTION

The characterization of material properties is critical to material and device design [1]. For the case of many magnetic materials this characterization is complicated by the need for materials with high refractive index for frequencies at or below L-band where their magnetic properties are most prominent. For these materials, free space techniques are impractical due to very large specimen lateral dimension requirements. As such, lumped circuit element (i.e., inductive and capacitive) fixtures, TEM transmission lines (e.g., coaxial airline and stripline), stripline resonant cavities, and rectangular waveguides are typically employed [2-7].

However, for materials with high refractive index, the permittivity and permeability errors are large for these fixtures. Specifically, stripline cavities suffer from cavity over perturbation, gap errors and uncertainty in specimen demagnetization factors. Lumped circuit element techniques suffer from gap errors and lumped circuit approximation errors as the fixture begins to demonstrate transmission line properties at UHF. Fully filled rectangular waveguides and coaxial airlines are plagued by gap errors. Motivated in large part by the desire to reduce the onerous specimen preparation requirements and to side step the error sources, research into the characterization of specimens which partially fill the cross section of rectangular waveguide (WG) has taken place in recent years [8-12]. The bulk of this research address only dielectric materials. References [13, 14] address magnetic materials but focus on higher frequencies (X-band and K-band) where magnetic properties are low.

The method presented in this work enables the permittivity and permeability characterization for specimens which possess (1) variable shape, (2) dimensions which need not fill either dimension of the WG cross section, and (3) have large permeability and permittivity values (>10). This is accomplished through the iterative comparison of simulations from a finite element model (FEM) of multiple waveguide-specimen geometries in a shorted rectangular WG with measurements in a like geometry. Two geometries are required to extract the unknown permittivity and permeability. Due to the use of the short in the WG fixture, the material parameter sensitivity of the measurement is improved relative to transmission and reflection measurements in an un-shorted WG. This permits the use of a specimen of minimum size for a given band, which can be utilized across multiple WG bands and thereby permits broadband characterization with a single specimen.

Material parameter extraction from the measured data relies on a sequential and iterative comparison of the measured and FEM-simulated scattering parameters ($S_{11}$) of the relevant WG-specimen geometries. During the frequency-by-frequency material parameter extraction, measured $S_{11}$ data for the requisite orientations are compared with many $S_{11}$ simulations obtained for various material parameter tensors. A population-based algorithm followed by a local gradient-based search controls the convergence of the extraction [15,16]. The specimen is assumed to obey effective media theory.

In order to demonstrate and validate the capability of the approach, permeability and permittivity values were obtained from measured $S_{11}$ data sets on two toroids of heavily loaded magnetic particle/polymer composites. A shorted WR1500 (500-750 MHz) WG measurement system was utilized. The accuracy of the results was supported, to the extent possible, through comparisons of extracted parameters with coaxial airline and stripline cavity measurements for the magnetic composite specimens.

## 2. PARAMETER EXTRACTION METHODOLOGY AND MEASUREMENT RESULTS

The extraction approach presented herein addresses isotropic complex material permittivity and permeability of the form $\varepsilon_r = \varepsilon'_r - j\varepsilon''_r$ and $\mu_r = \mu'_r - j\mu''_r$, respectively. The method obtains the individual values from measurements of specimens in a shorted rectangular WG and is accomplished in essentially three stages; (1) a preparation stage, (2) a measurement stage, and (3) an extraction stage. In each of the first two stages, all relevant dimensional information about the specimen, WG, and specimen location in the WG is collected. These data permit the extraction of the material parameters in the third stage through iterative comparison of simulated and measured *S*-parameters by facilitating the creation of the FEM model. The iteration is driven to convergence by a sequential global/local algorithm. Reference [15] details this process and a short summary will be repeated for this letter. The WG-specimen geometry shown in Fig. 1 will prove illuminating during the discussion which follows.

In order to probe all of the permittivity and permeability elements, a minimum of two measurements must be performed. For probing the permeability and permittivity, the specimen is placed flush against the short and a distance $l_{short}$ from the shorted end of the WG, respectively. To maximize sensitivity to the permittivity over the full band of the WG, $l_{short}$ is equal to the line standard length in the TRL calibration. In order to take advantage of both symmetry planes in the simulations, the specimen is always centered in the WG cross section. In order to further speed up simulations, network theory is used to obtain the *S* parameter results at the short position from the results at the offset position.

All measurement results presented herein were performed with an Agilent Network Analyzer and a shorted WR1500 WG measurement system, which was operated from 500-750 MHz. The shorted line was realized in the experiment by bolting a shorting plate to one end of the TRL calibration line standard. That ensemble was connected to the port one section of the waveguide system. The dashed line labeled *Short-Port 1 Interface* in

Fig. 1 (i.e., the $\hat{x}_{wg}$-$\hat{y}_{wg}$ plane at $z_{wg}=0$), represents this interface. The TRL calibrated measurements were further compensated by dividing the post-calibration specimen reflection measurements by measurements of the empty shorted section with the foam sample holder placed therein.

Material parameter extractions were performed on two material specimens. The two specimens were ARC-Technologies, Inc. UD-12300 and UD-115544 magnetic composites; the former being 0.241 cm thick and the latter being 0.444 cm thick. The specimens were manufactured with a 7.62 cm outer diameter and a 2.54 cm inner diameter. A photograph of the tested specimens is shown in Fig. 2.

Following the measurement details previously described and detailed in [15], $S_{11}$ versus frequency characteristics were obtained for the two WG-specimen geometries. Due to the commonality of the orientations at the short and offset positions, half the FEM simulations were required if coupled with the network theory analysis detailed in [7, 17]. The specimens were oriented in the waveguide such that the radial dimension was in the $\hat{x}_{wg}$-$\hat{y}_{wg}$ plane of the WG cross section. The specimens were each placed offset from and against the WG short yielding two geometries and one FEM simulation paired with network theory analysis. The dielectric specimen was positioned in the WG sequentially in four orthogonal orientations such that all permittivity components were probed. These orientations were realized offset from and against the short to probe the permeability, yielding eight orientations (two more than required) and four FEM simulations when paired with network theory analysis, illustrated in Fig. 3.

The extracted permittivity and permeability results for the magnetic polymer composite specimens are presented in Fig. 4(a) and Fig. 4(b), respectively. These extracted permittivity and permeability values were obtained by the method described in this letter. The extraction approach searched over a broad range of permittivity and permeability values. The upper and lower bounds for the inversion were 50 and 1 for the real parts of the permittivity and permeability respectively. Upper and lower bounds of 25 and 0 were used for the imaginary components. Simulation studies for the FEM meshes were performed at the upper bounds to assurance convergence.

Corroborating data, shown in Fig. 4, were obtained from additional measurements utilizing a stripline cavity for the permittivity data and a coaxial airline for the permeability data. Due to gap errors associated with coaxial airlines, the permittivity data were taken using the stripling cavity is lower than the partially filled waveguide data. Due to issues associated with demagnetization factors, the permeability data were taken from the coaxial airline. The coaxial data were provided by ARC-Technologies, Inc. It should be noted that the coaxial data was provided by the vendor and was obtained from a separate run of material. Comparisons of the measured results show reasonable agreement to better than 10% which is reasonable given the various potential error sources for the materials and fixtures.

## 3. CONCLUSIONS

A method for determining the permittivity and permeability for specimens with high material index and variable shape has been described. Representative measured results and material parameter extractions for magnetic materials have been presented and validated with corroborating measurements. The demonstrated method extracts the requisite material parameters from specimen reflection measurements made in a partially filled shorted rectangular waveguide. Measurements using WR1500 waveguide were made for two coaxial, heavily loaded magnetic composites. This work demonstrates an advancement in the state-of-the art of material characterization capability. Specifically, the first characterization of a UHF/VHF, high refractive index material utilizing a specimen which need not fill either dimension of the WG cross section (thereby avoiding gap errors). This approach further permits the flexibility to use an electrically small specimen of irregular shape across multiple waveguide bands and could be directly extended to characterize anisotropic material specimens.

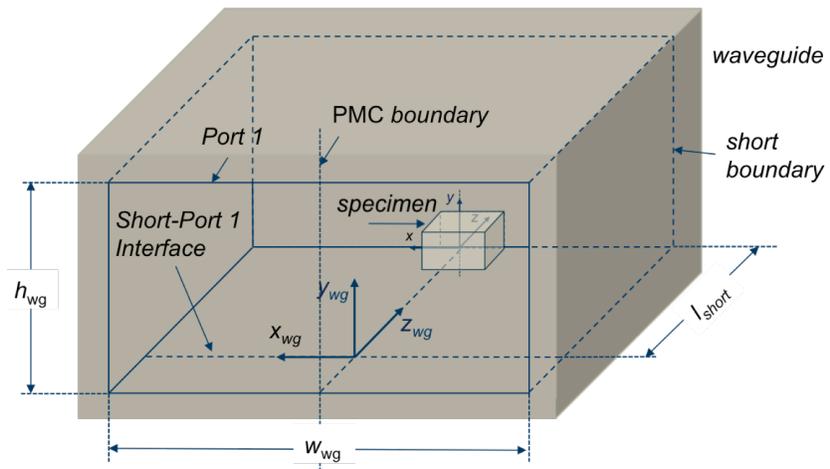

**Figure 1** Schematic representation of shorted waveguide measurement system and specimen.

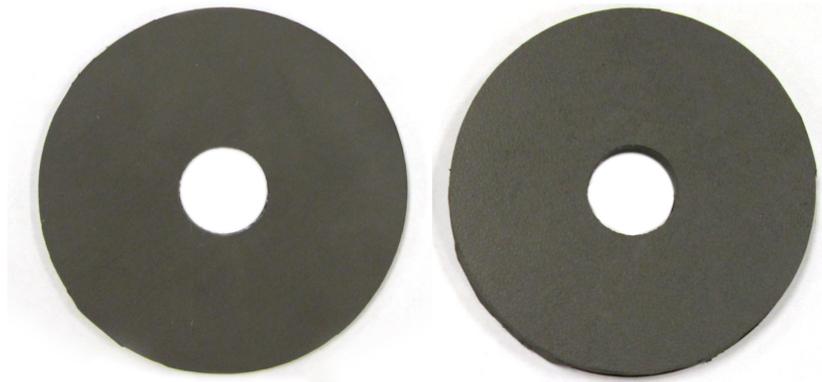

**Figure 2** Photograph of tested specimens, two 3 inch diameter magnetic polymer composites with 1 inch through-holes.

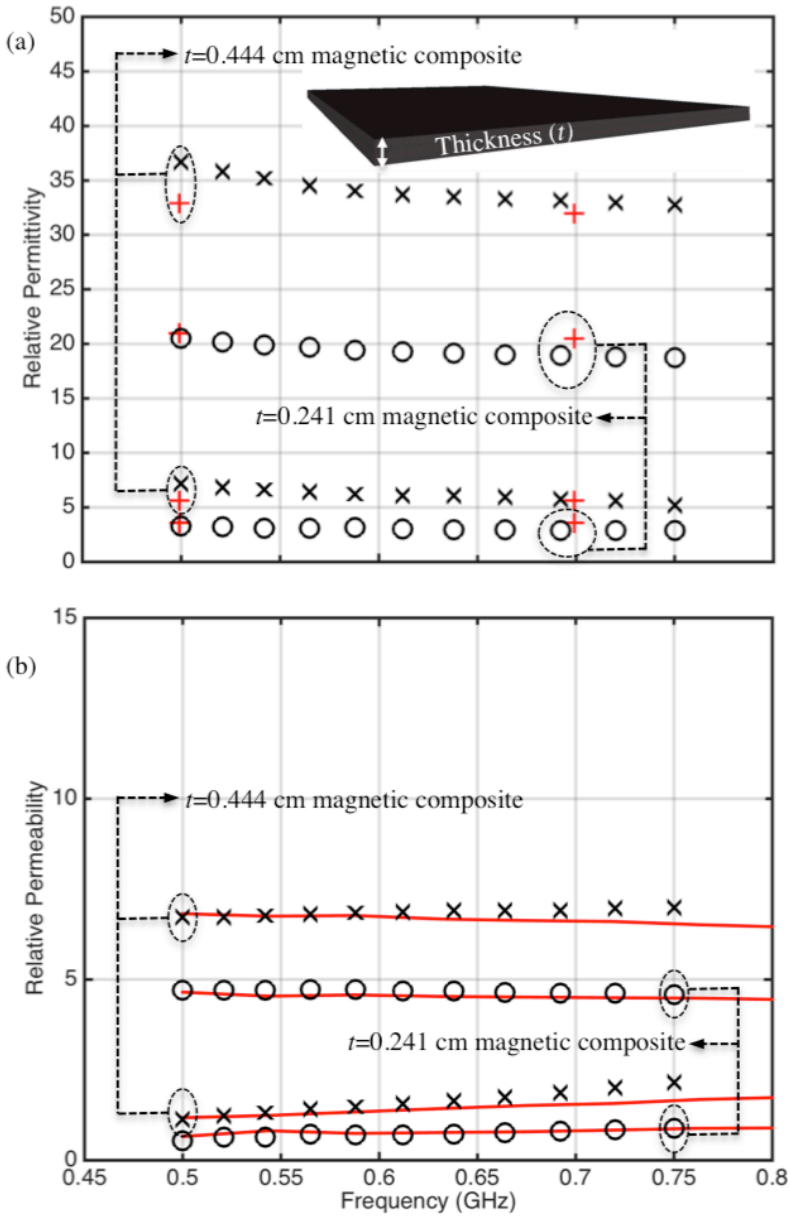

**Figure 3** Extracted (a) permittivity and (b) permeability for the 0.241 cm and 0.444 cm thick magnetic composite specimen represented by circles and x's, respectively. Supporting material parameter results from stripline cavity and coaxial airline are also shown as crosses and solid red lines, respectively.